\documentclass[aps,preprint,showkeys,showpacs]{revtex4}%
\usepackage{amsfonts}
\usepackage{amsmath}
\usepackage{amssymb}
\usepackage{graphicx}%
\begin{document}
\preprint{ }
\title{Inequalities for low-energy symmetric nuclear matter}
\author{Dean Lee}
\affiliation{North Carolina State University, Department of Physics, Raleigh, NC 27695, USA}

\begin{abstract}
Using effective field theory we prove inequalities for the correlations of
two-nucleon operators in low-energy symmetric nuclear matter. \ For physical
values of operator coefficients in the effective Lagrangian, the $S=1$, $I=0$
channel correlations must have the lowest energy and longest correlation
length in the two-nucleon sector. \ This result is valid at nonzero density
and temperature.

\end{abstract}
\maketitle

The study of inequalities in quantum chromodynamics (QCD) dates back over
twenty years. \ Even though the information one may extract from these
inequalities is limited, they provide rare insights into the nonperturbative
QCD vacuum \cite{Vafa:1983tf}\cite{Vafa:1984xg}\cite{Vafa:1984xh} and hadronic
spectrum \cite{Witten:1983ut}\cite{Nussinov:1983vh}\cite{Nussinov:2003uj}%
\cite{Cohen:2003ut} which currently cannot be addressed by other purely
analytic methods. Most rigorous QCD inequalities are derived using the
Euclidean functional integral. \ The close relationship to lattice QCD
simulations should be apparent. \ Typically one compares the magnitude of
various observable quantities by means of the Cauchy-Schwarz or H\"{o}lder
inequalities. \ The comparison requires positivity of the functional integral
measure for at least one of the quantities being compared and in some cases
positivity of the observable for arbitrary background gauge configurations.
\ These same attributes are also important in lattice QCD in connection with
the well-known sign/phase problem. \ A review of the literature on QCD
inequalities can be found in \cite{Nussinov:1999sx}.

Recently there has been some interest in simulations of nuclear matter on the
lattice using effective field theory \cite{Muller:1999cp}\cite{Chen:2003vy}%
\cite{Lee:2004si}\cite{Abe:2003fz}. \ These are low-energy simulations with
nucleons and sometimes pions as point particles rather than quarks and gluons.
\ Motivated by these nonperturbative studies, we consider here the subject of
effective field theory inequalities for nuclear matter. \ Our focus and
application will be to nuclear matter, but much of the interesting physics is
of a universal nature and appears in other systems such as trapped Fermi gases
near a Feshbach resonance \cite{Kinast:2004}\cite{Regal:2004}.

It should be obvious that effective field theory does not constrain properties
of the vacuum, nucleon and pion masses, or other few-body spectra at zero
chemical potential and temperature. \ These quantities are usually
phenomenological inputs which are used to determine the coefficients of
operators in the effective Lagrangian. \ Hence the inequalities which we prove
here concern correlation functions at nonzero chemical potential and/or
temperature. \ The results are most interesting at nucleon densities where the
average nucleon separation is less than the scattering length, and the physics
is strongly coupled.

We let $N$ represent the nucleon fields,%
\begin{equation}
N=\left[
\begin{array}
[c]{c}%
p\\
n
\end{array}
\right]  \otimes\left[
\begin{array}
[c]{c}%
\uparrow\\
\downarrow
\end{array}
\right]  . \label{nucfield}%
\end{equation}
We use $p$($n$) to represent protons(neutrons) and $\uparrow$($\downarrow$) to
represent up(down) spins. \ We use $\vec{\tau}$ to represent Pauli matrices
acting in isospin space and $\vec{\sigma}$ to represent Pauli matrices acting
in spin space. \ We assume exact isospin symmetry.\ \ In the non-relativistic
limit and below the threshold for pion production, we can write the lowest
order terms in the effective Lagrangian \cite{Weinberg:1990rz}%
\cite{Weinberg:1991um}\cite{Weinberg:1992yk}\cite{Kaplan:1996xu}%
\cite{Kaplan:1998tg}\cite{Kaplan:1998we} in several equivalent ways. \ Two of
the possibilities are%
\begin{equation}
\mathcal{L}=\bar{N}[i\partial_{0}+\tfrac{\vec{\nabla}^{2}}{2m_{N}}-(m_{N}%
^{0}-\mu)]N-\tfrac{1}{2}C_{S}\bar{N}N\bar{N}N-\tfrac{1}{2}C_{T}\bar{N}%
\vec{\sigma}N\cdot\bar{N}\vec{\sigma}N, \label{T}%
\end{equation}%
\begin{equation}
\mathcal{L}=\bar{N}[i\partial_{0}+\tfrac{\vec{\nabla}^{2}}{2m_{N}}%
-(m_{N}^{0\prime}-\mu)]N-\tfrac{1}{2}C_{S}^{\prime}\bar{N}N\bar{N}N-\tfrac
{1}{2}C_{U}^{\prime}\bar{N}\vec{\tau}N\cdot\bar{N}\vec{\tau}N. \label{U}%
\end{equation}
At this stage we leave out three-nucleon forces, but we will comment on this
later in our discussion. \ We assume that the high-momentum modes have been
removed by some regularization procedure though details are unimportant in
this discussion.

Consider states with two nucleons at the same point. \ In order to maintain
overall antisymmetry, we can have either a spin $S=0$, isospin $I=1$ state, or
an $S=1,I=0$ state. \ These two states are eigenstates of the various
interaction operators, and the corresponding eigenvalues are shown in Table 1.
\ At lowest order in the pionless effective theory, we do not have mixing
between $S$ and $D$ waves in the $S=1,I=0$ channel.%
\begin{equation}%
\genfrac{}{}{0pt}{0}{\text{Table 1: \ Eigenvalues of interaction operators}}{%
\begin{tabular}
[c]{|l|l|l|}\hline
& $S=1,I=0$ & $S=0,I=1$\\\hline
$\tfrac{1}{2}:\bar{N}N\bar{N}N:$ & $1$ & $1$\\\hline
$\tfrac{1}{2}:\bar{N}\vec{\sigma}N\cdot\bar{N}\vec{\sigma}N:$ & $1$ &
$-3$\\\hline
$\tfrac{1}{2}:\bar{N}\vec{\tau}N\cdot\bar{N}\vec{\tau}N:$ & $-3$ & $1$\\\hline
\end{tabular}
\ \ }%
\end{equation}
Clearly%
\begin{equation}
:\bar{N}N\bar{N}N:=-\tfrac{1}{2}:\bar{N}\vec{\sigma}N\cdot\bar{N}\vec{\sigma
}N:-\tfrac{1}{2}:\bar{N}\vec{\tau}N\cdot\bar{N}\vec{\tau}N:,
\end{equation}%
\begin{equation}
C_{U}^{\prime}=-C_{T},\quad C_{S}^{\prime}=C_{S}-2C_{T}.
\end{equation}
In the real world two-nucleon low-energy scattering is attractive in both the
$S=1,I=0$ and $S=0,I=1$ channels. \ It is also more strongly attractive in the
$S=1,I=0$ channel. \ From this we deduce that%
\begin{align}
C_{S}  &  <3C_{T},\quad C_{T}<0,\\
C_{S}^{\prime}  &  <-C_{U}^{\prime},\quad C_{U}^{\prime}>0.
\end{align}
These inequalities can be verified nonperturbatively by solving the
Schrodinger equation and determining phase shifts \cite{Lee:2004si} or by
summing all bubble diagrams \cite{Beane:2003da}. \ The analysis is
straightforward since we have only a simple contact potential.

The grand canonical partition function is given by%
\begin{equation}
Z_{G}\propto\int DND\bar{N}\exp\left(  -S_{E}\right)  =\int DND\bar{N}%
\exp\left(  \int d^{4}x\,\mathcal{L}_{E}\right)  ,
\end{equation}
where we choose to write $\mathcal{L}_{E}$ as%
\begin{equation}
\mathcal{L}_{E}=-\bar{N}[\partial_{4}-\tfrac{\vec{\nabla}^{2}}{2m_{N}}%
+(m_{N}^{0\prime}-\mu)]N-\tfrac{1}{2}C_{S}^{\prime}\bar{N}N\bar{N}N-\tfrac
{1}{2}C_{U}^{\prime}\bar{N}\vec{\tau}N\cdot\bar{N}\vec{\tau}N.
\end{equation}
Using Hubbard-Stratonovich transformations \cite{Stratonovich:1958}%
\cite{Hubbard:1959ub},\ we can rewrite $Z_{G}$ as%
\begin{equation}
Z_{G}\propto\int DND\bar{N}DfD\vec{\phi}\exp\left(  \int d^{4}x\,\mathcal{L}%
_{E}^{f,\vec{\phi}}\right)  ,
\end{equation}
where%
\begin{equation}
\mathcal{L}_{E}^{f,\vec{\phi}}=-\bar{N}[\partial_{4}-\tfrac{\vec{\nabla}^{2}%
}{2m_{N}}+(m_{N}^{0\prime}-\mu)]N+C_{S}^{\prime}f\bar{N}N+iC_{U}^{\prime}%
\vec{\phi}\cdot\bar{N}\vec{\tau}N-\mathcal{V}(f,\vec{\phi})
\end{equation}
and
\begin{equation}
-\mathcal{V}(f,\vec{\phi})=\tfrac{1}{2}C_{S}^{\prime}f^{2}-\tfrac{1}{2}%
C_{U}^{\prime}\vec{\phi}\cdot\vec{\phi}.
\end{equation}
We note that $C_{S}^{\prime}<0$ and $C_{U}^{\prime}>0,$ and so the $f$ and
$\vec{\phi}$ integrations are convergent.

Let $M$ be the nucleon matrix generated by the background Hubbard-Stratonovich
fields,%
\begin{equation}
M=-\left[  \partial_{4}-\tfrac{\vec{\nabla}^{2}}{2m_{N}}+(m_{N}^{0\prime}%
-\mu)\right]  +C_{S}^{\prime}f+iC_{U}^{\prime}\vec{\phi}\cdot\vec{\tau}.
\end{equation}
We note that%
\begin{equation}
\tau_{2}M\tau_{2}=M^{\ast}, \label{tau_2}%
\end{equation}
where $M^{\ast}$ is the complex conjugate of $M$, not the Hermitian adjoint.
\ We conclude that $\det M$ must be real. \ Since $M$ does not couple up and
down spins we can write $M$ with two identical diagonal blocks,%
\begin{equation}
M=%
\begin{bmatrix}
M_{\uparrow} & 0\\
0 & M_{\downarrow}%
\end{bmatrix}
,\quad M_{\uparrow}=M_{\downarrow}.
\end{equation}
We note that $\det M_{\uparrow}$ must also be real, and so
\begin{equation}
\det M=\det M_{\uparrow}\det M_{\downarrow}=(\det M_{\uparrow})^{2}\geq0.
\end{equation}
Although not relevant to the main line of this discussion, we point out that
actually $\det M_{\uparrow}\geq0$. \ This is because $\tau_{2}$ is
antisymmetric and therefore the real eigenvalues of $M_{\uparrow}$ are doubly
degenerate \footnote{Thanks to Jiunn-Wei Chen for discussions on this point.}.

We note the similarity between (\ref{tau_2}) and the relation%
\begin{equation}
\gamma_{5}M\gamma_{5}=M^{\dagger} \label{gamma5}%
\end{equation}
for the quark matrix in QCD at zero chemical potential. \ From (\ref{gamma5})
one can show positivity of the quark matrix determinant for two degenerate
flavors and the dominance of pion correlation functions
\cite{Weingarten:1983uj}. \ In our case, however, we observe that
(\ref{tau_2}) is also valid at nonzero chemical potential.

Let us now consider the two-nucleon operator%
\begin{equation}
A(x)=[N_{\uparrow}]_{i}[\tau_{2}]_{ij}[N_{\downarrow}]_{j}(x)
\end{equation}
where $N_{\uparrow}$ and $N_{\downarrow}$ are up and down spin projections of
$N$. \ Since $A$ has quantum number $I=0,$ overall antisymmetry requires it
also have quantum number $S=1.$ \ The two-point correlation function for $A$
is%
\begin{equation}
\ \left\langle A(x)A^{\dagger}(0)\right\rangle _{\mu,T}=\left\langle
[N_{\uparrow}]_{i}[\tau_{2}]_{ij}[N_{\downarrow}]_{j}(x)\;[N_{\downarrow
}^{\ast}]_{l}[\tau_{2}]_{lm}[N_{\uparrow}^{\ast}]_{m}(0)\right\rangle _{\mu
,T}.
\end{equation}
Using our Euclidean functional integral representation, we have%
\begin{equation}
\left\langle A(x)A^{\dagger}(0)\right\rangle _{\mu,T}=\int D\Theta
\;[M_{\downarrow}^{-1}(x,0)]_{jl}[\tau_{2}]_{lm}[M_{\uparrow}^{-1}%
(x,0)]_{im}[\tau_{2}]_{ij},
\end{equation}
where $D\Theta$ is the positive normalized measure defined by%
\begin{equation}
D\Theta=\dfrac{DfD\vec{\phi}\;\det M\exp\left(  -\int d^{4}x\,\mathcal{V}%
(f,\vec{\phi})\right)  }{\int DfD\vec{\phi}\;\det M\exp\left(  -\int
d^{4}x\,\mathcal{V}(f,\vec{\phi})\right)  }.
\end{equation}
From the symmetry properties of $\tau_{2}$ under transposition and
(\ref{tau_2}),%
\begin{align}
\lbrack M_{\downarrow}^{-1}(x,0)]_{jl}[\tau_{2}]_{lm}[M_{\uparrow}%
^{-1}(x,0)]_{im}[\tau_{2}]_{ij}  &  =[M_{\downarrow}^{-1}(x,0)]_{jl}[\tau
_{2}]_{ji}[M_{\uparrow}^{-1}(x,0)]_{im}[\tau_{2}]_{ml}\nonumber\\
&  =[M_{\downarrow}^{-1}(x,0)]_{jl}[M_{\uparrow}^{-1}(x,0)]_{jl}^{\ast
}\nonumber\\
&  =\sum_{jl}\left\vert [M_{\uparrow}^{-1}(x,0)]_{jl}\right\vert ^{2}.
\end{align}
Thus%
\begin{equation}
\left\langle A(x)A^{\dagger}(0)\right\rangle _{\mu,T}=\int D\Theta\sum
_{jl}\left\vert [M_{\uparrow}^{-1}(x,0)]_{jl}\right\vert ^{2}.
\end{equation}

Consider now the more general operator%
\begin{equation}
B(x)=[N_{\uparrow}]_{i}[b]_{ij}[N_{\downarrow}]_{j}(x)\label{specific}%
\end{equation}
for arbitrary $2\times2$ matrix $[b]_{ij}$. \ We find
\begin{equation}
\left\langle B(x)B^{\dagger}(0)\right\rangle _{\mu,T}=\int D\Theta G_{B}(x),
\end{equation}
where%
\begin{equation}
G_{B}(x)=[M_{\downarrow}^{-1}(x,0)]_{jl}[b]_{ml}^{\ast}[M_{\uparrow}%
^{-1}(x,0)]_{im}[b]_{ij}.
\end{equation}
From the Cauchy-Schwarz inequality,%
\begin{align}
\left\vert G_{B}(x)\right\vert  &  \leq\sqrt{\sum_{ijlm}\left\vert
[M_{\uparrow}^{-1}(x,0)]_{im}\right\vert ^{2}\left\vert [M_{\downarrow}%
^{-1}(x,0)]_{jl}\right\vert ^{2}}\sqrt{\sum_{i^{\prime}j^{\prime}l^{\prime
}m^{\prime}}\left\vert [b]_{i^{\prime}j^{\prime}}\right\vert ^{2}\left\vert
[b]_{m^{\prime}l^{\prime}}\right\vert ^{2}}\nonumber\\
&  \leq\sum_{im}\left\vert [M_{\uparrow}^{-1}(x,0)]_{im}\right\vert ^{2}%
\sum_{jk}\left\vert [b]_{jk}\right\vert ^{2}.
\end{align}
We conclude that%
\begin{equation}
\left\vert \left\langle B(x)B^{\dagger}(0)\right\rangle _{\mu,T}\right\vert
\leq\int D\Theta\left\vert G_{B}(x)\right\vert \leq\gamma\left\langle
A(x)A^{\dagger}(0)\right\rangle _{\mu,T}\label{main}%
\end{equation}
where
\begin{equation}
\gamma=\sum_{jk}\left\vert [b]_{jk}\right\vert ^{2}.\label{constant}%
\end{equation}

Let us take the zero temperature limit at fixed nonzero nucleon density.
\ This will require tuning the chemical potential $\mu$ to match the desired
density as we lower the temperature. If we take the separation $x\rightarrow
\infty$ in the Euclidean time direction, we deduce from (\ref{main}) that the
energy of the lowest state with the quantum numbers of $A$ (i.e., $S=1$,
$I=0$) must be less than or equal to the lowest state with the quantum numbers
of $B$,%
\begin{equation}
E_{A}^{0}\leq E_{B}^{0}. \label{energy}%
\end{equation}
\ This result was known from the outset for zero density. \ The non-trivial
result is that it also holds true at nonzero density. \ In particular if we
are in a region of the phase diagram where the deuteron is unbound, then all
states with the quantum numbers of $B$ must also be unbound. \ A formula due
to L\"{u}scher \cite{Luscher:1986pf} tells us that in a periodic box of length
$L$, the energy of the lowest two-particle scattering state relative to
threshold is%
\begin{equation}
E_{0}=\dfrac{4\pi a}{m_{N}L^{3}}[1+O(aL^{-1})], \label{lus}%
\end{equation}
where is $a$ is the scattering length with sign convention,%
\begin{equation}
k\cot\delta_{0}=-\frac{1}{a}+\frac{1}{2}r_{0}k^{2}+....
\end{equation}
By combining (\ref{energy}) and (\ref{lus}), we find in the unbound case the
following inequality for the scattering lengths \footnote{Thanks to Thomas
Schaefer for suggesting the application of L\"{u}scher's formula for the
unbound case.},%

\begin{equation}
a_{A}\leq a_{B}. \label{scatt}%
\end{equation}

If we take the limit $x\rightarrow\infty$ in any spatial direction, we find
that the correlation length for $A$ must be greater than or equal to the
correlation length for $B$,%
\begin{equation}
\xi_{A}\geq\xi_{B}.\label{length}%
\end{equation}
This holds true for any density and any temperature. \ In particular if
$\xi_{B}$ diverges, as would occur in a superfluid phase transition, then
$\xi_{A}$ must also diverge. \ In that case one can use (\ref{main}) and
(\ref{constant}) to bound the relative sizes of the dinucleon condensates.

The inequalities (\ref{energy}), (\ref{scatt}), and (\ref{length}) were proven
for operators $B$ of the specific form given in (\ref{specific}). \ It is
straightforward to generalize to the case when the nucleons have arbitrary
spin directions and/or are not at the same point. Let%
\begin{equation}
B_{b,y}(x)=[N(x+y)]_{i}[b]_{ij}[N(x-y)]_{j},
\end{equation}
where $[b]_{ij}$ is a general $4\times4$ matrix. \ One can show for some
positive parameters $\gamma_{1}$, $\gamma_{2}$, $\gamma_{3}$, $\gamma_{4}$,%
\begin{align}
\left\vert \left\langle B_{b^{\prime},y^{\prime}}(x)B_{b,y}^{\dagger
}(0)\right\rangle _{\mu,T}\right\vert  &  \leq\gamma_{1}\left\langle
A(x-y^{\prime})A^{\dagger}(-y)\right\rangle +\gamma_{2}\left\langle
A(x-y^{\prime})A^{\dagger}(y)\right\rangle \nonumber\\
&  +\gamma_{3}\left\langle A(x+y^{\prime})A^{\dagger}(-y)\right\rangle
+\gamma_{4}\left\langle A(x+y^{\prime})A^{\dagger}(y)\right\rangle . \label{B}%
\end{align}
From (\ref{B}) we conclude that the inequalities (\ref{energy}),
(\ref{scatt}), and (\ref{length}) hold for any two-nucleon operator $B$,%
\begin{equation}
B(x)=\int_{\Omega}d^{4}y\;[N(x+y)]_{i}[b(y)]_{ij}[N(x-y)]_{j},
\end{equation}
where the region of integration $\Omega$ is bounded.

In summary we have shown that the $S=1$, $I=0$ channel must have the lowest
energy and longest correlation length in the two-nucleon sector. \ These
results are valid at arbitrary density and temperature and can be readily
checked in lattice simulations using chiral effective field theory. \ Assuming
that the effective Lagrangian can describe the relevant physics, it would be
interesting to study two-nucleon correlations near the region of the phase
diagram where the deuteron becomes unbound. \ In principle these results can
also be checked in lattice QCD at finite density, though the complex phase
problem will make the calculation exceedingly difficult.

It turns out that the addition of the three-nucleon force term $(\bar{N}%
N)^{3}$ \cite{Bedaque:1999ve} can be introduced by modifying the coupling of
$f$ and $\bar{N}N$\ while preserving positivity of $\det M$ and the conjugate
similarity relation in (\ref{tau_2}). \ Other modifications can also be made
for various higher order terms in the effective Lagrangian. \ These and other
extensions will be discussed in future work.

The author thanks Paulo Bedaque, Jiunn-Wei Chen, and Thomas Schaefer for
several helpful discussions. \ He also thanks Paulo for organizing the Summer
of Lattice Workshop 2004 at Lawrence Berkeley Laboratory where part of this
research was completed. \ This work was supported by the Department of Energy.

\bibliographystyle{h-physrev3}
\bibliography{NuclearMatter}

\begin{thebibliography}{10}

\bibitem{Vafa:1983tf}
C.~Vafa and E.~Witten,
\newblock Nucl. Phys. {\bf B234}, 173 (1984).

\bibitem{Vafa:1984xg}
C.~Vafa and E.~Witten,
\newblock Phys. Rev. Lett. {\bf 53}, 535 (1984).

\bibitem{Vafa:1984xh}
C.~Vafa and E.~Witten,
\newblock Commun. Math. Phys. {\bf 95}, 257 (1984).

\bibitem{Witten:1983ut}
E.~Witten,
\newblock Phys. Rev. Lett. {\bf 51}, 2351 (1983).

\bibitem{Nussinov:1983vh}
S.~Nussinov,
\newblock Phys. Rev. Lett. {\bf 51}, 2081 (1983).

\bibitem{Nussinov:2003uj}
S.~Nussinov,
\newblock (2003), hep-ph/0306187.

\bibitem{Cohen:2003ut}
T.~D. Cohen,
\newblock Phys. Rev. Lett. {\bf 91}, 032002 (2003), hep-ph/0304024.

\bibitem{Nussinov:1999sx}
S.~Nussinov and M.~A. Lampert,
\newblock Phys. Rept. {\bf 362}, 193 (2002), hep-ph/9911532.

\bibitem{Muller:1999cp}
H.~M. M{\"u}ller, S.~E. Koonin, R.~Seki, and U.~van Kolck,
\newblock Phys. Rev. {\bf C61}, 044320 (2000), nucl-th/9910038.

\bibitem{Chen:2003vy}
J.-W. Chen and D.~B. Kaplan,
\newblock (2003), hep-lat/0308016.

\bibitem{Lee:2004si}
D.~Lee, B.~Borasoy, and T.~Schafer,
\newblock (2004), nucl-th/0402072.

\bibitem{Abe:2003fz}
T.~Abe, R.~Seki, and A.~N. Kocharian,
\newblock (2003), nucl-th/0312125.

\bibitem{Kinast:2004}
J.~Kinast, S.~L. Hemmer, M.~E. Gehm, A.~Turlapov, and J.~E. Thomas,
\newblock Phys. Rev. Lett. {\bf 92}, 150402 (2004).

\bibitem{Regal:2004}
C.~A. Regal, M.~Greiner, and D.~S. Jin,
\newblock Phys. Rev. Lett. {\bf 92}, 040403 (2004).

\bibitem{Weinberg:1990rz}
S.~Weinberg,
\newblock Phys. Lett. {\bf B251}, 288 (1990).

\bibitem{Weinberg:1991um}
S.~Weinberg,
\newblock Nucl. Phys. {\bf B363}, 3 (1991).

\bibitem{Weinberg:1992yk}
S.~Weinberg,
\newblock Phys. Lett. {\bf B295}, 114 (1992), hep-ph/9209257.

\bibitem{Kaplan:1996xu}
D.~B. Kaplan, M.~J. Savage, and M.~B. Wise,
\newblock Nucl. Phys. {\bf B478}, 629 (1996), nucl-th/9605002.

\bibitem{Kaplan:1998tg}
D.~B. Kaplan, M.~J. Savage, and M.~B. Wise,
\newblock Phys. Lett. {\bf B424}, 390 (1998), nucl-th/9801034.

\bibitem{Kaplan:1998we}
D.~B. Kaplan, M.~J. Savage, and M.~B. Wise,
\newblock Nucl. Phys. {\bf B534}, 329 (1998), nucl-th/9802075.

\bibitem{Beane:2003da}
S.~R. Beane, P.~F. Bedaque, A.~Parreno, and M.~J. Savage,
\newblock (2003), hep-lat/0312004.

\bibitem{Stratonovich:1958}
R.~L. Stratonovich,
\newblock Soviet Phys. Doklady {\bf 2}, 416 (1958).

\bibitem{Hubbard:1959ub}
J.~Hubbard,
\newblock Phys. Rev. Lett. {\bf 3}, 77 (1959).

\bibitem{Weingarten:1983uj}
D.~Weingarten,
\newblock Phys. Rev. Lett. {\bf 51}, 1830 (1983).

\bibitem{Luscher:1986pf}
M.~L{\"u}scher,
\newblock Commun. Math. Phys. {\bf 105}, 153 (1986).

\bibitem{Bedaque:1999ve}
P.~F. Bedaque, H.~W. Hammer, and U.~van Kolck,
\newblock Nucl. Phys. {\bf A676}, 357 (2000), nucl-th/9906032.

\end{thebibliography}

\end{document}